\documentclass[11pt]{article}

\usepackage{fullpage}
\usepackage{amsmath}
\usepackage{amssymb}
\usepackage{mathrsfs}
\usepackage{setspace}
\usepackage{bbm}
\usepackage{dsfont}

\usepackage{graphics}

\usepackage[latin1]{inputenc}
\usepackage[pdftex]{graphicx}

\usepackage{color}
\usepackage[american]{babel}

\newcommand{\be}{\begin{equation}}
\newcommand{\ee}{\end{equation}}
\newcommand{\bea}{\begin{eqnarray}}
\newcommand{\eea}{\end{eqnarray}}

\newcommand{\p}{\partial}

\newcommand{\mat}{\left(\begin{array}{cc}}
\newcommand{\emat}{\end{array}\right)}
\newcommand{\smat}{\left(\begin{smallmatrix}}
\newcommand{\esmat}{\end{smallmatrix}\right)}

\newcommand{\R}{\mathbb{R}}
\newcommand{\Z}{\mathbb{Z}}

\begin{document}
\numberwithin{equation}{section}
{
\begin{titlepage}
\begin{center}

\hfill \\
\hfill \\
\vskip 0.75in

{\Large \bf Comments on AdS$_2$ Gravity}\\

\vskip 0.4in

{Alejandra Castro${}^{a}$ and  Wei Song${}^{b,c}$}\\

\vskip 0.3in
{\it ${}^a$ Institute for Theoretical Physics, University of Amsterdam,
Science Park 904, Postbus 94485, 1090 GL Amsterdam, The Netherlands }
\vskip .5mm 
 {\it ${}^b$ Mathematical Sciences Center, Tsinghua University, Beijing, 100084,China }
 \vskip .5mm 
 {\it ${}^c$ Department of Physics, Princeton University, Princeton, NJ 08544, USA}
\vskip 0.3in

\end{center}

\vskip 0.35in

\begin{abstract} 
\noindent
We revisit some properties of AdS$_2$ Einstein-Maxwell gravity  with the aim of reconciling apparently conflicting results in prior literature. We show that the two dimensional theory can be obtained as a dimensional reduction of the three dimensional Einstein gravity with the CSS boundary condition \cite{Compere:2013bya}. In general, this theory on AdS$_2$  can be viewed as an effective theory of the near horizon geometry of a (near-)extremal black hole. We provide an interpretation of our results in terms of the microscopic description of an extremal black hole. 
\end{abstract}

\vfill


\end{titlepage}
}

\newpage
\tableofcontents


\section{Introduction}

AdS$_2$ is a generic (and natural) fixed point for a wide variety of gravitational systems.   It is the attractor geometry of a large class of extremal black holes, and for this reason,  the dual theory  we assign to AdS$_2$  will correspond to a universal IR fixed point of the UV theory that describes these black holes. However, AdS$_2$ gravity is a rather enigmatic subject. In the context of holography, the two disconnected boundaries of the spacetime is alone enough evidence to argue that new qualitative features are need. It is as well bizarre that the ground state of this theory carries a large amount of entropy: it is difficult to derive such a large degeneracy in the dual field theory without the aid of, for example, supersymmetry. 

Our goal is to infer universal properties of the theory dual to AdS$_2$. In particular we would like to tie  the analysis of \cite{Hartman:2008dq,Castro:2008ms} for AdS$_2$ black holes with the recent results in AdS$_3$ \cite{Compere:2013bya}.  Another important portion of our analysis is to interpret the structures found in AdS$_2$ in the context of the near horizon physics of a higher dimensional extremal black hole.

The analysis of asymptotic symmetries is a systematic method to extract properties of the candidate dual theory.  
In the context of AdS$_3$ gravity this has been widely successful. The original work of Brown-Henneaux \cite{Brown:1986nw} showed that under certain boundary conditions, the asymptotic symmetry group for AdS$_3$ in Einstein gravity is generated by left and right moving Virasoro algebras.
This is strong evidence that  quantum gravity on AdS$_3$ is holographically dual to a two dimensional conformal field theory. Furthermore, the Bekenstein-Hawking entropy for BTZ black holes is reproduced by the Cardy formula of the dual CFT$_2$ \cite{Strominger:1997eq}. 

However, a generic extremal black hole has fewer global symmetries than those preserved by Brown-Henneaux boundary conditions. We are aiming to describe a system  with a global  $ SL(2,\mathbb{R})_R\times U(1)_L$ which are the isometries of the near horizon geometry.  One way to reduce the number of symmetries is by imposing alternative boundary conditions for AdS$_3$; these have been recently discussed \cite{Compere:2013bya, Troessaert:2013fma}. In CSS \cite{Compere:2013bya}, parity violating boundary conditions are chosen, so that the $SL(2,\mathbb{R})_L\times SL(2,\mathbb{R})_R $ isometry of AdS$_3$ are only enhanced to a right-moving Viraroso-Kac-Moody algebra, with central charge $c_R$ and level $k$. A field theory with such a Virasoro-Kac-Moody structure is known as Warped Conformal Field Theory (WCFT) \cite{Hofman:2011zj,Detournay:2012pc,Hofman:2014loa}.\footnote{Chiral Liouville gravity  \cite{Compere:2013aya} is an explicit example of WCFT; examples involving Weyl fermions are given in \cite{Hofman:2014loa}. } Despite some odd features, which we discuss in section \ref{sec:pi}, the degeneracy of states of a WCFT still captures the Bekenstein-Hawking entropy of BTZ \cite{Detournay:2012pc}. 

An alternative (and more general) way to reduce the isometries is to simply consider different backgrounds which do not have an AdS$_3$ origin. This is the approach taken in, e.g. \cite{Compere:2008cv,Compere:2014bia} which considered warped AdS$_3$ black holes, and in \cite{Guica:2008mu} where the NHEK geometry (near horizon of extremal Kerr black holes) was studied. One important feature of these backgrounds is that the local isometry contains $SL(2,\mathbb{R})_R\times U(1)_L$. 
A direct asymptotic symmetry analysis \cite{Compere:2008cv} leads to the right-moving Virasoro-Kac-Moody algebra, implying the dual theory is a WCFT instead of an ordinary CFT. However, a different construction of the asymptotic symmetric analysis, which uses the symplectic structure, leads to a left and right moving Virasoro algebra \cite{Compere:2014bia}. For warped AdS$_3$ black holes, the Bekenstein-Hawking entropy can be either reproduced by a Cardy formula for an ordinary CFT \cite{Anninos:2008fx}, or a WCFT \cite{Detournay:2012pc}.  At this stage it is not clear which description (CFT or WCFT) is more natural or which one admits a UV description. We also note that warped AdS$_3$ can be also viewed as a continuous deformation from ordinary AdS$_3$. It turns out the continuous limit of the natural boundary conditions for warped AdS$_3$ is the boundary conditions in \cite{Compere:2013bya}, instead of the Brown-Henneaux boundary condition.

To summarize, there is some evidence indicating that a possible description of a  geometry with $SL(2,\mathbb{R})_R\times U(1)_L$ isometries is a WCFT.   Since the Virasoroso-Kac-Moody for WCFT is purely right moving, one might retain certain features of the algebra upon dimensional  reduction along the $U(1)_L$ direction. The resulting theory will be  a two dimensional gravitational theory, where  AdS$_2$ with a background $U(1) $ gauge field is a solution. One question we would like to address is: {\it does AdS$_2$ quantum gravity know about WCFTs?}

With this motivation, we will revisit the issue of boundary conditions for AdS$_2$ in Maxwell-Dilaton gravity and try to reconcile and clarify some conflicting results among \cite{Hartman:2008dq,Castro:2008ms,Compere:2013bya}. We will show how AdS$_2$ boundary conditions in \cite{Castro:2008ms} are compatible with the new boundary conditions in three dimensions \cite{Compere:2013bya}. In comparison with  \cite{Hartman:2008dq},  their is an apparent conflict with \cite{Compere:2013bya} which arises because their current algebra does not transform under the improved stress tensor canonically. In  appendix \ref{app:hs}, we review the derivation in the bulk approach of \cite{Hartman:2008dq} and show that if requiring the canonical algebra between the stress tensor and U(1) current, the stress tensor should be further improved. The resulting relation between the level and central charge agrees with our approach.

The layout of this paper is as follows. In section 2, we reduce the new boundary conditions  for AdS$_3$ and interpret them in AdS$_2$. In section 3, we uplift the two dimensional theory to five dimensions; here we provide evidence that CSS boundary conditions are those that naturally allow for superradiant modes. We end up with some discussions  and open directions in section 4.

\section{From AdS$_3$ to AdS$_2$}\label{sec:pi}

In this section we will review the main results of \cite{Compere:2013bya}, and show how it has an extremely natural interpretation in AdS$_2$ upon reduction.  Our starting point is pure AdS$_3$ gravity; the action is given by 
\be\label{act3d}
I_{3D}={1\over 16\pi G_3}\int d^3x \sqrt{-g^{(3)}}\left(R^{(3)}+{2\over \ell^2}\right)~,
\ee
and we consider a class of solutions with the following asymptotics 
\bea\label{eq:ads3}
{ds^2_{3D}\over \ell^2}&=& {dr^2\over r^2} - r^2(dx^+dx^--\partial_+P(x^+)(dx^+)^2)\cr
&&+ {4G_3\over \ell}\Delta\left(dx^--\partial_+P(x^+)dx^+ \right)^2 +{4 G_3\over \ell} L(x^+) (dx^+)^2  +O(r^{-2})~.
\eea
Here $x^\pm=t\pm \phi$ with $\phi\sim \phi +2\pi$; $\Delta$ is a constant, while $P(x^+)$ and $L(x^+)$ are arbitrary functions of $x^+$. In this notation, the BTZ black hole with mass $M$ and angular momentum $J$ corresponds to 
\be
L(x^+)=L~,\quad \partial_+P(x^+)=0~,\quad \ell M=\Delta+L~,\quad \ell J=\Delta-L~,
\ee
 where $L$ is  constant and $\Delta>0$. Global AdS corresponds to $\Delta=L=-{\ell/ G_3}$ and  $\partial_+P(x^+)=0$.

The allowed fluctuations of the metric under diffeomorphism are  
\be\label{eq:allowed3d}
\delta g_{++}= O(r^2) ~, \quad \delta g_{+-} = O(1)~,\quad \delta g_{--}=O(r^{-2})~,
\ee
and the choice of radial gauge is fixed as well. In other words, the allowed diffeomorphism leave the leading term of $g_{+-}$ and $\Delta$  fixed as well; the functions $P(x^+)$ and $L(x^+)$ are allowed to fluctuate. Solving for \eqref{eq:allowed3d} gives rise to an asymptotic symmetry group generated by the vectors
\bea\label{diffasg}
\epsilon&=& \zeta(x^+)\partial_+ - {r\over 2}\partial_+ \zeta(x^+)\partial_r +\cdots~,\cr
\eta &=& \mu(x^+)\partial_-+\cdots~.
\eea
Note that if we do not allow fluctuations of $\partial_+P(x^+)$, this analysis will boil down to the holomorphic sector of the Brown-Henneaux boundary conditions -- these are the boundary conditions considered in e.g. \cite{Gupta:2008ki,Balasubramanian:2009bg}.   

The canonical charges associated to the generators \eqref{diffasg} are \cite{Compere:2013bya}
\bea\label{eq:3q}
Q_{\zeta}&=&{1\over 2\pi} \int_0^{2\pi}d\phi\, \zeta(x^+)\left[L(x^+)-\Delta (\partial_+ P(x^+))^2\right]~,\cr
Q_{\mu}&=&{1\over 2\pi} \int_0^{2\pi}d\phi\, \mu(x^+)\left[\Delta +2\Delta \partial_+ P(x^+)\right]~.
\eea
Setting $\zeta=\zeta_n:= e^{inx^+}$ and $\mu=\mu_n:=e^{inx^+}$, the algebra of charges is
\bea\label{alg3d}
i\{Q_{\zeta_n},Q_{\zeta_m}\}&=&(n-m)Q_{\zeta_{n+m}}+{c\over12} m(m^2-1)\delta_{n+m}~,\cr
i\{Q_{\zeta_n},Q_{\mu_m}\}&=&-m Q_{\mu_{n+m}}~,\cr
i\{Q_{\mu_n},Q_{\mu_m}\}&=&{k\over 2}m\delta_{{n+m}}~,
\eea
which corresponds to a Virasoro coupled to a Kac-Moody algebra and the central extensions are  
\be\label{eq:ck3}
c={3\ell\over 2G_3}~,\quad k=-4\Delta~.
\ee
This algebra is the defining feature of WCFT \cite{Hofman:2011zj,Detournay:2012pc,Compere:2013aya}. They are characterized by having only one copy of the Virasoro algebra, and a $U(1)$ Kac-Moody which is not an internal symmetry.\footnote{The current algebra here describes diffeomorphisms. Furthermore it is in the same sector as the Virasoro algebra. We emphasize that \eqref{alg3d} is not a truncation of the two copies of the Virasoro algebra obtained by the usual Brown-Henneaux boundary conditions. } The rather novel property here is that the level $k$ depends on the zero mode of $Q_{\mu_n}$, i.e. on the value of $\Delta$. In particular for the black hole background, where $\Delta>0$, the level $k$ is negative which makes the theory non-unitary.\footnote{To properly discuss unitarity we need to define as well an inner product. Here we are assuming usual Hermitian conditions, i.e. $Q^\dagger_n = Q_{-n}$. However there could be a different choice that allows for unitary representations of \eqref{alg3d}. } 

Despite the apparent lack of unitarity of the AdS$_3$ boundary conditions proposed by CSS \cite{Compere:2013bya} we would like to present how they appear universally in AdS$_2$; in the following section we will  give a physical interpretation of such behavior. To make contact with AdS$_2$, we Kaluza-Klein reduce the Lagrangian \eqref{act3d}, where we  decompose the three dimensional metric as
\be\label{kkmetric}
ds_{3D}^2= e^{-2\psi} \ell^2(dz+A_\mu dx^\mu)^2+g_{\mu\nu}dx^\mu dx^\nu ~,\quad \mu=0,1~.
\ee
We will take the fields $\psi$, $A_\mu$ and $g_{\mu\nu}$ to be independent of the $z$ direction, and we will compactify  along  $z\sim z+2\pi$. Replacing in \eqref{act3d}, up to total derivatives, we find the 2D action 
\be\label{2dactionx}
I_{2D}={\ell\over 8 G_3}\int d^2x \sqrt{-g}e^{-\psi}\left(R^{(2)}+{2\over \ell^2}-{\ell^2\over 4}e^{-2\psi}F^{2}\right)~,
\ee
where we used
\be
R^{(3)}=R^{(2)}-2e^{-\psi}\nabla^2 e^{-\psi}-{\ell^2\over 4} e^{-2\psi} F^2 ~,\quad F_{\mu\nu}=\partial_\mu A_\nu -\partial_\nu A_\mu~.
\ee
 The two and three dimensional couplings  are related via
\be\label{eq:g3g2}
{ G_3}= { 2 \pi \ell G_2}~. 
\ee
%

Next, lets discuss briefly the discussion of asymptotic symmetries from the 2D perspective. This theory admit solutions that are locally AdS$_2$ provided that $\psi$ is constant. An asymptotically AdS$_2$ has the following  Fefferman-Graham expansion \cite{Castro:2008ms}
\be
{ds^2_{\rm 2D }\over \ell^2}={d\sigma^2\over 4\sigma^2} + h_{tt} dt^2~,\quad A =A_t dt~,
\ee
where
\bea\label{bcond}
h_{tt}&=& h^{(0)} \sigma^{-2}  + h^{(2)} +  O(\sigma^{2})~,\cr
A_t&=& A^{(0)} \sigma^{-1} + A^{(2)} + A^{(4)}\sigma + O(\sigma^{2})~,\cr
e^{-\psi}&=&e^{-\psi^{(0)}}~.
\eea
The parameters $h^{(2)}$,  $A^{(2)}$ and $A^{(4)}$ are  functions of $t$. The dependence on the radial variable $\sigma$ is constructed by requiring compatible with the equations of motion.
Still not all parameters in \eqref{bcond} are independent. We leave the details for appendix \ref{sec:ads2}; the bottom line is that equations of motion constrain certain components of the fields. In particular, after imposing \eqref{eom1}  we find at leading order
\be\label{eombndy}
h^{(0)} + e^{-2\psi^{(0)}} A^{(0)} A^{(0)} =0~, 
\ee
and for the subleading term
\be\label{ln:eom}
h^{(2)}= 2A^{(0)}A^{(4)} e^{-2\psi^{(0)}}~.
\ee

It is useful to lift the boundary conditions for AdS$_2$; replacing the Fefferman-Graham  expansion \eqref{bcond} in \eqref{kkmetric} gives 
\bea\label{dd:ads2}
{ds^2_{3D}\over \ell^2}&=& {d\sigma^2\over 4\sigma^2} +  \sigma^{-2}\left(h^{(0)} +  e^{-2\psi^{(0)}} A^{(0)} A^{(0)}\right)dt^2  + {2\over \sigma} \left(A^{(0)}dzdt+A^{(0)}A^{(2)}dt^2\right)e^{-2\psi^{(0)}}\cr
&&+  e^{-2\psi^{(0)}} \left(dz+A^{(2)}dt\right)^2 +\left(h^{(2)}+2A^{(0)}A^{(4)}e^{-2\psi^{(0)}}\right)dt^2  +O(\sigma)~.
\eea
Using the on-shell  condition \eqref{eombndy} the leading boundary term cancels; the boundary metric then is $\sim dzdt$. Hence the coordinates $(z,t)$ are naturally null coordinates in AdS$_3$ and the Kaluza-Klein reduction is along one of the null $z$  direction. The sign of $e^{-2\psi^{(0)}}$ determines whether  the reduction is timelike, null or spacelike. 
Relabeling  the coordinates in \eqref{dd:ads2} as
\be
r^2=2A^{(0)}e^{-2\psi^{(0)}}\sigma^{-1}~,\quad z=-x^-~,\quad t= x^+~,
\ee
 then \eqref{dd:ads2} reads
\bea\label{eq:ads3}
{ds^2_{3D}\over \ell^2}&=& {dr^2\over r^2} - r^2(dx^+dx^--A^{(2)}(dx^+)^2)\cr
&&+  e^{-2\psi^{(0)}} \left(dx^--A^{(2)}dx^+ \right)^2 +2 h^{(2)} (dx^+)^2  +O(r^{-2})~,
\eea
where we made use of \eqref{eombndy} and \eqref{ln:eom}.   At this stage we are pleased to find a exact match of the  boundary conditions \eqref{eq:allowed3d} with \eqref{bcond}: in term of the variables used there we have
\be\label{eq:vv}
 A^{(2)}(x^+)=\partial_+\bar P ~,\quad \ell e^{-2\psi^{(0)}}=4G_3\Delta ~,\quad h^{(2)}(x^+) =2G_3\ell \bar L(x^+)~.
\ee

In the two dimensional theory, the boundary conditions \eqref{bcond} are preserved by a set of  diffeomorphism and $U(1)$ gauge transformations \cite{Hartman:2008dq}; we review this construction in appendix \ref{sec:bndy}. These set of transformations   map precisely to the  diffeomorphisms \eqref{diffasg}: $\epsilon$ implements ``diff+gauge'' in AdS$_2$ and $\eta$ is a residual gauge transformation of $A_\mu$.

Next we should discuss the role of the algebra \eqref{alg3d} from the AdS$_2$ perspective. A first comparison is with the work of Hartman and Strominger in \cite{Hartman:2008dq}. An interpretation of their result is that  provided there exists a stress and a $U(1)$ current  along the lines of  \eqref{alg3d}, the relative sign between the central charge and level will be negative in accordance with \eqref{eq:ck3}.\footnote{This derivation is reviewed and clarified in appendix \ref{app:hs}.} Combined with our above observations, this motivates us to propose that a dual theory of AdS$_2$ is not necessarily the chiral half of a 2D CFT (or a DLCQ of it), but rather it could be as well a WCFT.

One might expect to make a more quantitative comparison of either the charges or the central charges between two and three dimensions. Unfortunately, a direct  comparison of the 3D charges \eqref{eq:3q} with a counterpart derivation in 2D is rather ambiguous. In appendix \ref{sec:charge} we discuss in detail what are the obstructions behind a construction of charges in AdS$_2$. At most we could suggest a candidate stress tensor ($\cal T$) and a current ($J_t$) in AdS$_2$ which have anomalous transformations. There are many weak points of this derivation, which are explained in appendix \ref{sec:charge}. Still it is interesting to note that if we compare the anomalies of these densities in \eqref{eq:ck2d} with those in \eqref{eq:ck3} we would find
\bea
k = -4\Delta =  -{e^{-2\psi^{(0)}}\over 2\pi G_2}=  k_{\rm 2D}~, \cr
c= {3\ell\over 2 G_3} = {3\over 4\pi G_2}= c_{\rm 2D}~.
\eea
It would be interesting to improve (and justify) the derivations in appendix \ref{sec:charge}, and hence make this comparison robust. Regardless this failure on our part, it should be clear that the construction of ``new boundary conditions'' for AdS$_3$ done in \cite{Compere:2013bya} is actually the natural set of boundary conditions used in the holographic description of AdS$_2$ Einstein-Maxwell gravity. This provides a more broad picture of what are possible descriptions of the dual theory for AdS$_2$ which we'll discuss in the following section. 

Finally, it is interesting to notice is that in the AdS$_3$ analysis the vacuum expectation value of the scalar field $\psi$ (or equivalently $\Delta$)  is not necessarily  positive. For the purposes of the Kaluza-Klein reduction to AdS$_2$ we restrict the discussion to $e^{2\psi^{(0)}}>0$, which further implies that $k$ is strictly negative. The effective 2D theory does not capture the other branches  of solutions with $\Delta\leq0$ which are natural in the 3D theory (e.g. global AdS$_3$ has $\Delta =-1$).

\section{Higher dimensional perspective: gluing negative energy photons}\label{sec:glue}

The near horizon geometry of a vast class of extremal black holes has an AdS$_2$ factor \cite{Kunduri:2007vf,Kunduri:2013gce}, i.e. the isometries of  the near horizon geometry contain at least an $SL(2,\R)$ factor. We can regard the theory on the near horizon background as a two dimensional theory obtained as a result of compactifying the additional directions.  A sector of the resulting theory will be in the same universality class as the action \eqref{2dactionx} where the higher dimensional origin of Maxwell field could be, for example, a non-trivial $S^1$ fibration over AdS$_2$ due to rotation---and of course there will be additional matter fields to those presented in \eqref{2dactionx}.  The  features presented in the previous section could be applicable to any of these solutions; in this section we want to argue when it is natural to describe the dual of an extremal black hole in terms of a WCFT.

The effective theory on AdS$_2$---with our choice of boundary conditions---has the peculiar feature that for positive values of the central charge $c_{\rm 2D}$, the level $k_{\rm 2D}$ is negative.\footnote{A naive construction of the Hilbert space for this algebra would indicate that there are negative norm states.  Similar behavior has been encountered in other holographic settings \cite{Castro:2012bc,Afshar:2012hc}.}  The ``offensive,'' and apparently worrisome, relation can be tracked back to the mode $A^{(2)}$ in \eqref{bcond} (or $\partial_+ P$ in 3D variables). This mode has as well a negative contribution to the total energy of the system \eqref{eq:3q}.  However, we will argue in this section that  the actual physical interpretation of these states relies as well on boundary conditions at asymptotic infinity. An apparent instability of the near horizon region does not necessarily trigger an instability of the full spacetime. We will consider a very explicit example and illustrate how global conditions on the geometry will give a natural interpretation to the $A^{(2)}$ modes.

To be concise, we  will consider the extremal black holes solutions studied in \cite{Song:2011ii}.\footnote{See as well \cite{Bena:2012wc} for further examples and generalizations.} The appeal of these examples is that  a 5D extremal spinning black hole shares the same near horizon geometry as a black string in Taub-Nut space.  This feature will allow us to clearly  illustrate the role that global properties of the black hole and black string have in giving an interpretation to the near horizon states.

It is easier to carry out the computations if we lift the solutions to 6D as done in \cite{Song:2011ii}; and  for the purpose of our argument we only need to focus on the metric. Further details of the full solution, including matter fields, can be found in \cite{Song:2011ii} and references within. The six dimensional near horizon region of interest is
\bea\label{eq:nhglue}
{ds^2\over \ell^2}&=&{d\sigma^2\over \sigma^2}-{dt^2\over \sigma^2}+\gamma(dy+\sigma^{-1} dt)^2+\gamma(d\psi +\cos\theta d\phi)^2\cr&&+2\alpha\gamma(dy+\sigma^{-1} dt)(d\psi+\cos\theta d\phi)+d\theta^2+\sin^2\theta d\phi^2~,
\eea
with
\be
y\sim y+4\pi^2 T_R m ~,\quad \psi\sim \psi+2\pi \Theta m +4\pi n ~,\quad m,n \in \Z~.
\ee
Here the constants $(\ell,\alpha,\gamma)$ and the periodicities $(T_R,\Theta)$ will be functions of the conserved charges of black hole (or string) the solution.  The AdS$_2$ portion of the geometry is describe by the $(\sigma,t)$ coordinates. We have two gauge fields arising from the cross terms in the metric in $(t,y)$ and $(t,\psi)$ directions. Shifts of the form
\be\label{eq:a2s}
\psi ~\to ~\psi +A_\psi^{(2)} t ~,\quad y~\to ~y +A_y^{(2)} t ~,
\ee
are the boundary transformations that generate the $U(1)$ Kac-Moody states.


There are (at least) two possible gluing of this near horizon metric to an asymptotic far region, i.e. two distinct extremal solutions who share the near horizon geometry \eqref{eq:nhglue}. To start, we first connect the near horizon region \eqref{eq:nhglue} to black hole with a far region given by $\R^{4,1}\times S^1$. Following the notation in  \cite{Song:2011ii}, the metric of extremal spinning black hole is
\bea\label{eq:mpmetric}
ds^2&=&-{(\hat r^2 +a^2(1-4c^2))\over \Sigma} d\hat t^2 +d\hat u^2+{4a^2\sinh2\delta\over \Sigma}d\hat t d\hat u + \Sigma{\hat r^2 d\hat r^2\over (\hat r^2 -a^2)^2}\cr&& +{\Sigma \over 4}\left(d\theta^2 + \sin\theta^2 d\phi^2\right) + \left({\Sigma\over 4}+{a^4\over \Sigma}\right)(d\hat\psi +\cos\theta d\hat \phi)^2\cr &&-{4a^3\over \Sigma}((c^3+s^3)d\hat t +sc(s+c)d\hat u)(d\hat\psi +\cos\theta d\hat \phi)~,
\eea
 where we introduced the notation
 \be
 \Sigma= \hat r^2 +a^2(1+4s^2) ~, \quad c=\cosh\delta ~,\quad s=\sinh\delta~.
 \ee
The solution carries mass, angular momentum and electric charge, and its relation to the parameters $a$ and $\delta$ is given by
\bea
M=6a^2\cosh 2\delta~,\quad
J_L=4a^3(c^3+s^3)~,\quad
Q=4a^2 sc~.
\eea
The asymptotic far region, $\hat r\to \infty$, is simply $\R^{4,1}\times S^1$:
\bea\label{farMP}
ds^2 \underset{\hat r\to \infty }\rightarrow -d\hat t^2 +d\hat u^2+d\hat r^2 +{\hat r^2 \over 4}\left(d\theta^2 + \sin\theta^2 d\phi^2+ (d\hat\psi +\cos\theta d\hat \phi)^2\right)\eea
where
\be
\hat u\sim \hat u +2\pi R m ~,\quad \hat\psi \sim \hat \psi + 4\pi n~,\quad m,n \in \Z~,
\ee
and $R$ is the size of the $S^1$ circle. Finally notice that the metric \eqref{eq:mpmetric} has an ergoregion.\footnote{ It is simple to check that any timelike Killing vector at infinity becomes spacelike before reaching the black hole horizon; this defines the presence of an ergoregion.}

To obtain the near horizon geometry we  define
\bea\label{eq:ppp}
&&\hat r^2={\epsilon\over \sigma} +a^2~,\quad \hat t={2a^2\over \Omega_L}{t\over \epsilon}~,
\cr
&&\hat u={1\over 2\pi T_Q}\left(y+\chi {t\over \epsilon}\right)~,\quad 
\hat \psi =\psi +{ 2a^2}{t\over \epsilon}+{J_L\Omega_L\over 4a^2}y ~,
\eea
where
\be
T_Q={c^3-s^3\over 4\pi a s^2 c^2}~ ,\quad \Omega_L={1\over a(c^3-s^3)} ~,\quad \chi= {a\over \Omega_L}{s-c\over sc}~.
\ee
The near  horizon limit is given by taking $\epsilon\to 0$ while the un-hatted variables are fixed. This gives precisely the line element \eqref{eq:nhglue} where we identify
\be
\ell^2={M\over 12}~,\quad \alpha={2\cosh2\delta\over 1+\cosh^22\delta}~,\quad \gamma=1+{1\over \cosh^22\delta}~,
\ee
and the periodicities of the $(y,\psi)$ directions are
\be
\Theta= -{2J_L\over Q^2}R~,\quad T_R=RT_Q~.
\ee

The coordinate transformations for $(\hat \psi,\hat u)$ in \eqref{eq:ppp} are of the form \eqref{eq:a2s}, which indicates that the $A^{(2)}$ modes are used in matching the near horizon region to the full solution. So let's interpret the negative energy photons from the perspective of the asymptotic region \eqref{farMP}.  In the near horizon region we  measured energies with respect to $\partial_t$, whereas in the full spacetime  we would use $\partial_{\hat t}$. From \eqref{eq:ppp} they are related as
\bea\label{eq:nhmyp}
\epsilon \partial_t &=& {2 a^2\over \Omega_L} \partial_{\hat t} + { 2a^2} \partial_{\hat \psi} + {\chi\over 2\pi T_Q}\partial_{\hat u} \cr 
&\equiv&{2 a^2\over \Omega_L} \partial_{\hat t} + A_\psi^{(2)} \partial_{\hat \psi} + A_u^{(2)}\partial_{\hat u} ~,
\eea
where in the second line we identified the shifts in \eqref{eq:ppp} as the boundary photons  in \eqref{eq:a2s}. 
From here we see that  this relation implies 
\be\label{mmpp}
\epsilon^2 |\partial_t |^2 \underset{\hat r\to \infty }\to   \hat r^2 (A_\psi^{(2)})^2 ~,
\ee
where we used \eqref{farMP}.
The vector $\partial_{t}$, which is timelike with respect to the AdS$_2$ metric,  becomes spacelike at infinity. The vector $\partial_{\hat t}$ has the opposite behavior: it is timelike at $\hat r\to \infty$ and spacelike near $\hat r\to a$. The change in sign of the norm of  $\partial_t$ and $\partial_{\hat t}$ occurs precisely when the ergosphere is crossed. 

We see that a positive energy mode at infinity can become negative near the horizon (and vice versa). The rather natural reason for this is the presence of superradiance scattering in the black hole background; and as it is well known this is due to the presence of an ergoregion in the full geometry. Hence we can interpret the seemingly offensive  Kac-Moody states in the near horizon geometry as a trigger to superradiance instability. 

The shifts \eqref{eq:ppp} ensures that the desired asymptotic behavior \eqref{farMP} is obtained;  it is not a harmless (i.e. pure gauge) transformation  due to the radial dependence $g_{\hat \psi\hat \psi}$ at large $\hat r$.  In contrast, notice that $A_u^{(2)}$ is not important for this argument: these photons remain pure gauge due to the simple fact that the size of the $S^1$ remains constant as we glue the near horizon geometry \eqref{eq:nhglue} to $\R^{4,1}\times S^1$.

Alternatively, we can glue these photons to an asymptotic region in a different manner, and in particular consider a gluing that does not generate an ergosphere. For this purpose we consider a black string in a Taub-Nut background. The six dimensional metric is given by
\bea\label{eq:bss}
ds^2&=&-H^{-2}d\hat t^2+H^2(d\hat r ^2 +\hat r^2(d\theta^2 +\sin^2\theta d\phi^2)) \cr&&+P^2\gamma_e \left[(d\hat \psi +\cos\theta d\phi+\alpha_e A_e)^2+(d\hat u +\sqrt{1-\alpha_e^2}A_e)^2\right] ~,
\eea
where
\be
H=1+{P\over \hat r}~,\quad A_e=-{d\hat t \over \hat r +P}~.
\ee
The constants $(P,\alpha_e,\gamma_e)$ determine the various RR and KK charges carried by the solution, see \cite{Song:2011ii} for explicit expressions. At large $\hat r$, we have
\be\label{farS}
ds^2 \underset{\hat r\to \infty }\to -d\hat t^2 +d\hat r^2 +{\hat r^2}\left(d\theta^2 + \sin\theta^2 d\phi^2\right)+ P^2\gamma_e(d\hat\psi +\cos\theta d\hat \phi)^2 + P^2\gamma_ed\hat u^2~.
\ee
The asymptotic region is locally $\R^{3,1}\times S^1\times S^1$ and we assign periodicities
\be
\hat u\sim \hat u +2\pi \hat R\, m~,\quad \hat \psi\sim \hat \psi +2\pi  \hat \Theta m+4\pi n~,\quad m,n \in \Z~.
\ee
The string is compactified along $\hat u$ with KK radius $\hat R$, and has twisted boundary conditions along $\hat \psi$. In contrast to the extremal black hole \eqref{eq:mpmetric}, the black string lacks an ergoregion. 

The near horizon of the black string is precisely given by \eqref{eq:nhglue}. More explicitly, re-define the coordinates as
\bea\label{eq:bbb}
&&\hat r=\epsilon r~, \quad \hat t=P^2{t\over \epsilon}~,\cr
&&\hat u=\sqrt{1-\alpha_e^2}\left(y+P{t\over \epsilon}\right)~,\quad
\hat \psi =\psi +\alpha_e\left(y+P{t\over \epsilon}\right)~.
\eea
The limit $\epsilon\to 0$ then gives the line element \eqref{eq:nhglue}, where we identify
\be
\ell^2= P^2~,\quad\alpha_e= \alpha ~,\quad \gamma_e=\gamma~,
\ee
and
\be
T_R= {\hat R\over 2\pi \sqrt{1-\alpha_e^2}}~,\quad \Theta=\hat \Theta- {\hat R\alpha_e\over  \sqrt{1-\alpha_e^2}}~. 
\ee

Next we interpret the near horizon photons from the point of view of the far region for the black string.  As for the spinning black hole, the transformations \eqref{eq:bbb} involve shifts of $(\hat \psi,\hat u)$ of the form \eqref{eq:a2s}. This gives a relation between energy states near the horizon and infinity:   
\bea
\epsilon \partial_t = P^2 \partial_{\hat t} + \sqrt{1-\alpha_e^2} P\partial_{\hat u} +\alpha_e P \partial_{\hat \psi} ~.
\eea
Using \eqref{farS} we find
\be
\epsilon^2 |\partial_t |^2 \underset{\hat r\to \infty }\to P^4(-1+\gamma_e) ~.
\ee
Provided that $\gamma_e<1$, the positive energy states in the near horizon region remain positive as viewed from the far region. In sharp contrast to the situation in \eqref{mmpp}, here negative norm states in the interior remain unphysical as we reach the asymptotic boundary. Furthermore, notice that for the black string both $S^1$ directions remain unaltered as we interpolate between far and near region. The fact that $g_{\hat \psi \hat \psi}$ and $g_{\hat u \hat u}$ are independent of the radial direction implies in this case an absence of ergoregion; furthermore it implies that the boundary photons remain pure gauge modes.  

We conclude that for the black string the negative norm photons are spurious states which should be excluded from the spectrum. On the other hand, for the black hole these states are very much physical and in part describe the superradiance instability inherent of spinning black holes. 

\section{Discussion}

 We provided some evidence towards the dual description of the near horizon of extremal black holes in terms of a WCFT. In AdS$_3$ gravity, the warped theory has the feature that the central charge and level have opposite signs \cite{Compere:2013bya}. This is shown explicit for a broad class of AdS$_2$ theories in appendix  \ref{app:hs} following \cite{Hartman:2008dq}.

To understand the interpretation of the warped boundary conditions, in section \ref{sec:glue} we interpolated the AdS$_2$ geometry to an asymptotic space where  the UV theory is better understood. In this setup we linked the existence of the negative norm states to superradiant excitations of the full solution. This gives a more direct and physical motivation to consider the new boundary conditions of AdS$_3$ presented in \cite{Compere:2013bya} in the context of near horizon geometries of extremal black holes. To support this interpretation, it would be interesting to actually implement the boundary conditions of \cite{Compere:2013bya} in \eqref{eq:nhglue}, and moreover cast the amplification of the modes due to superradiant scattering in the language of a WCFT. We leave this analysis for future work.

Our analysis in AdS$_2$ seems to not distinguish between supersymmetric and non-supersymmetric black holes, but of course there is a distinction! The UV theory (far region) dictates which are the appropriate boundary conditions that define the IR theory (near region). The example  in section \ref{sec:glue} reflects this choice. It might be possible that non-supersymmetric black holes are better described by  a WCFT, while supersymmetric black holes follow the  successful framework built in \cite{Sen:2008yk,Sen:2008vm}.\footnote{For a non-supersymmetric black hole it is not clear to us if the concept of quantum entropy function is well defined, and hence which aspects in \cite{Sen:2008vm} will apply to any extremal black hole is not obvious.} The fact that the WCFT is not unitary could simply indicate that the IR system is not fully decoupled from its parent UV theory. Still this possibility needs to be investigated further. 

The example in section \ref{sec:glue} attempts to give guidance for when a WCFT is an appropriate dual. However, there are still cases where there is room for ambiguity. For example, a BPS string corresponds to the limiting case of the solution \eqref{eq:bss} with $\gamma_e=1$;  in this limit the AdS$_2$ fibration \eqref{eq:mpmetric} enhances to AdS$_3$. This case is rather special. From \eqref{eq:ads3} the fiber can be either null or timelike, i.e.  $e^{2\psi^{(0)}}\leq0$, and this changes the physical interpretation of \eqref{eq:ck3}. But most importantly, there is still room for two very distinct boundary conditions: one which allows $A^{(2)}$ to  fluctuate and have a theory with one copy of Virasoro plus a Kac-Mooody algebra, or fix $A^{(2)}$ which reduces to a chiral sector of a 2D CFT along the lines of  e.g. \cite{Gupta:2008ki,Balasubramanian:2009bg}. These are two very different theories, with different physical implications for the black hole. Understanding when either boundary condition is the ``correct one''\footnote{We emphasize that there are several cases where fixing $A^{(2)}$ is the correct picture for supersymmetric black holes as it has been shown explicitly with high precision; see  \cite{Sen:2008vm, Mandal:2010cj} for a full set of references. The question we are raising is if there exist a string theory setup where the boundary conditions presented here and in \cite{Compere:2013bya} are the relevant one to describe the IR physics. }  is an extremely interesting question which we will pursue in future work.

Even though we did not construct examples where the full black hole solution is asymptotically AdS, our analysis should carry through smoothly. We expect that the negative norm photons constructed here can be interpreted as well as superradiance instabilities.  Instabilities of the AdS$_2$ geometry have been explored in for both AdS$_4$ and AdS$_5$ black holes (see e.g. \cite{Almuhairi:2011ws,Donos:2011qt,Donos:2011ff} and references within);  it would be interesting to relate these instabilities and understand more sharply the properties of the dual theory.

To end we should mention how prior work and our analysis is related to the original Kerr/CFT conjecture \cite{Guica:2008mu}. There, the analysis was based on a boundary condition which enhances $U(1)_L$ to a left-moving Virasoro. There are so far two pillars supporting an ordinary CFT. The first pillar is that the Cardy's formula reproduces the Bekenstein-Hawking entropy \cite{Guica:2008mu}, and the second pillar is that the two point function reproduces the greybody factor in the bulk \cite{Bredberg:2009pv,Cvetic:2009jn}. One obstacle against an ordinary CFT is that so far there are no consistent boundary conditions such that both the left and right moving conformal symmetries can be found simultaneously, although some effects have been made along this direction \cite{Matsuo:2009sj,Matsuo:2010ut,Guica:2011ia,Guica:2013jza,Compere:2014bia}.  An alternative proposal is that the dual theory is not an ordinary CFT, but instead a WCFT.\footnote{From the field theory side, a closely related question is what the local symmetry  for  a two dimensional field theory with global translation invariance plus a right-moving scaling invariance. It was shown in \cite{Hofman:2011zj} that under certain reasonable assumptions, the right-moving scaling symmetry is always enhanced to conformal symmetry. For the left-moving sector, there are two-minimal options. one possibility is that the left-moving sector also has conformal symmetry. The resulting theory will be an ordinary two dimensional CFT. Another possibility is that the left-moving translation symmetry is enhanced to a right-moving Kac-Moody algebra, leading to a WCFT.} If the second picture is realized, there must be some boundary condition such that the asymptotic symmetry group of NHEK is a pure right-moving Virasoro-Kac Moody. This are still open questions that cannot fully answer yet but we hope our present analysis will elucidate what is the correct path.

\section*{Acknowledgements}

We thank St\'ephane Detournay, Tom Hartman, Diego Hofman, Andrew Strominger for useful conversations, and Finn Larsen for encouragement. We specially thank Andrew Strominger for collaborations at the early stage of this work; we are grateful of Geoffrey Compère and Tom Hartman for helpful comments on the draft. A.C. thanks the organizers and participants of ``Holography for Black Holes and Cosmology Workshop'' at the Solvay Institute, and  ``Black Objects Beyond Supersymmetry Workshop'' at Universiteit Utrecht. A.C.'s work was partially supported in part by the Fundamental Laws Initiative of the Center for the Fundamental Laws of Nature, Harvard University, and  by Nederlandse Organisatie voor
Wetenschappelijk Onderzoek (NWO) via a Vidi grant.   W.S. was supported in part by the Harvard Society of Fellows, by the U.S. Department of Energy under the grant number DE-FG02-91ER40671, and by the start-up funding 543310007 from Tsinghua University.

\appendix

\section{AdS$_2$ Maxwell-Dilaton Gravity}\label{sec:ads2}

 In this appendix will review some aspects of 2D gravity. We will mostly focus on the results of  \cite{Grumiller:2007ju, Hartman:2008dq,Castro:2009jf} which have a clear interpretation in AdS$_3$; towards the end of the appendix we will comment on some puzzles which we couldn't resolve. 

There are many versions of 2D gravity that admit AdS$_2$ as a solution. Here we will work with the following Maxwell-Dilaton action
\be\label{2daction}
I_{\rm 2D}={1\over  16 \pi G_2}\int d^2x \sqrt{-g}e^{-\psi}\left(R^{(2)}+{2\over \ell^2}-{\ell^2\over 4}e^{-2\psi}F^{2}\right)~.
\ee
The precise couplings we introduced in \eqref{2daction} are simply a convenient choice that will allow us to compare directly with 3D. As it will become evident below, the key ingredients in our analysis is the presence of a $U(1)$ gauge field and the specification of boundary conditions.  In the following subsections we will review the construction of the variation principle and asymptotic symmetry group following   \cite{Castro:2008ms}; see as well   \cite{Grumiller:2007ju, Hartman:2008dq,Castro:2009jf}.

\subsection{Equations of motion}\label{sec:eom}

 Our focus will be on solutions with constant $\psi$;\footnote{It would be interesting to consider configurations with non-constant dilaton and its possible interpretations. See \cite{Grumiller:2013swa} for a recent discussion.} in this  case the equations of motion reduce to 
 \be\label{eom1}
 R^{(2)}=-{8\over \ell^2}~,\quad e^{-2\psi} F^2=-{8\over \ell^4}~,\quad e^{-3\psi}  \partial_\mu\left(\sqrt{-g}F^{\mu\nu}\right)=0~.
 \ee
Next, to characterize the solutions,  we pick a  gauge where
\be\label{gauge}
{ds^2\over \ell^2}={d\sigma^2\over 4\sigma^2} + h_{tt} dt^2~,\quad A_\sigma =0~.
\ee
Imposing the equations of motion gives 
\bea\label{eq:eom11}
h_{tt}&=& -e^{-2\psi}\left({(A^{(0)})^2\over\sigma^2}-2A^{(0)}A^{(4)}+(A^{(4)})^2\sigma^2\right)\cr
 A_t &=&{A^{(0)}\over\sigma} + A^{(2)}+A^{(4)}\sigma,
\eea
where $A^{(0)},\,A^{(2)}$ and $A^{(4)}$  are arbitrary functions of $t$. We can always set $h^{(0)}=-e^{-2\psi}(A^{(0)})^2$ to be a constant by re-definition of the time coordinate, and $A^{(2)}$ is seemingly pure gauge.

\subsection{Boundary conditions and asymptotic symmetries}\label{sec:bndy}


We are interested in classical configurations that asymptote to AdS$_2$. More concretely, at the boundary ($\sigma\to 0$) we want our fields to have the local isometries of $sl(2,\R)$ group of diffeomorphism. In terms of the fields \eqref{gauge}, we are considering classical solution that asymptote to   
\bea\label{bb:bc}
h_{tt}|_{\sigma\to0}&=&h^{(0)} \sigma^{-2}~, \quad A_t|_{\sigma\to0}= A^{(0)} \sigma^{-1}~,\quad e^\psi|_{\sigma\to0}=e^{\psi^{(0)}}~,
\eea
where $h^{(0)}$, $A^{(0)}$ and $e^{\psi^{(0)}}$ are constant.

Given these boundary conditions we will now construct the set of 
diffeomorphisms and $U(1)$ gauge transformations that preserve these boundary conditions. For the metric, we want to find the set of diffeomorphisms $x^\mu\to x^\mu +\epsilon^\mu$ that satisfy\footnote{In our conventions we have $
\delta_\epsilon g_{\mu\nu}=-\nabla_\mu \epsilon_\nu - \nabla_\nu \epsilon_\mu
 $.}
\bea\label{bc1}
\delta_{\epsilon} g_{\sigma\sigma}= 0 ~,\quad  \delta_{\epsilon} g_{\sigma t}= 0~,\quad \delta_{\epsilon} g_{tt}=O(1)~,
\eea
i.e. that the gauge choice \eqref{gauge} is preserved, and $h^{(0)}$ is fixed while $h^{(2)}$ is allowed to fluctuate.  The diffeomorphism that preserve   \eqref{bc1}  are
\bea\label{2d:diffasg}
\epsilon^t&=& \zeta(t) -{1\over 8 h^{(0)}}(1-{\sigma^2h^{(2)}\over2h^{(0)}})\partial_t^2 \zeta(t)\sigma^2+\cdots~,\cr
\epsilon^\sigma &=& \sigma\partial_t \zeta(t)~.
\eea
Here $\zeta(t)$ is an arbitrary function of $t$. Under these vectors, the metric transforms as
\be\label{eq:he}
\delta_\epsilon h_{tt}= -\zeta \partial_t h^{(2)} - 2h^{(2)}  \partial_t \zeta +{1\over 4} \partial_t^3\zeta+\cdots~.
\ee

We also have to preserve the boundary conditions for the gauge field, i.e.
\be\label{bcgauge}
 \delta_{\epsilon +\Lambda} A_\sigma =0 ~,\quad \delta_{\epsilon +\Lambda} A_t = O(1)~,
\ee
where we include both coordinate and gauge transformations:
\bea
\delta_{\epsilon +\Lambda}A_\mu := -(\epsilon^{\lambda}\nabla_\lambda A_\mu+A_{\lambda}\nabla_\mu \epsilon^\lambda)+\partial_\mu \Lambda~.
\eea
Under the set of  diffeomorphisms \eqref{2d:diffasg},  we find
\be
\delta_\epsilon A_\sigma = -{A^{(0)}\over \sigma}\partial_\sigma \epsilon^t=  {A^{(0)}\over 4 h^{(0)}}\partial_t^2 \zeta(t)~,
\ee
and hence does not preserve the boundary conditions in accordance to the observation made in \cite{Hartman:2008dq}. To compensate for the non-trivial action of the diffeomorphism, and  preserve \eqref{bcgauge}, we act as well with a gauge transformation described by 
\be\label{gaugeasg}
\Lambda = -{1\over 4 h^{(0)}}\partial_t^2\zeta\left(A^{(0)}\sigma+{1\over 2}A^{(2)} \sigma^2+\cdots\right)~,
\ee
and hence we can set $\delta_{\epsilon+\Lambda} A_\sigma=0$ order by order in $\sigma$. Under the combined diffeomorphisms \eqref{2d:diffasg} and gauge parameter \eqref{gaugeasg} the time component of the gauge field transforms as 
\be\label{eq:AAA}
\delta_{\epsilon+\Lambda} A_t= -\zeta \partial_t A^{(2)} -A^{(2)}  \partial_t \zeta-\sigma\left( \zeta \partial_t A^{(4)} + 2A^{(4)}  \partial_t \zeta +{A^{(0)}\over 8h^{(0)}} \partial_t^3\zeta\right)+\cdots~,
\ee
which preserves the boundary condition \eqref{bcgauge}. 

Finally, there is one additional transformation which preserves our boundary conditions. This is simply given by $U(1)$ gauge transformation $\Lambda=\mu(t)$
\be\label{eq:simu1}
\delta_{\mu} A_t = \partial_t \mu  ~,\quad \delta_{\mu} A_\sigma = 0~,
\ee
which only modifies the $A^{(2)}$ component of the Fefferman-Graham expansion \eqref{bcond}.

Summarizing, the asymptotic symmetries of the theory are generated by the diffeomorphisms \eqref{2d:diffasg}, which {\it must} be accompanied by the gauge transformation \eqref{gaugeasg}, and the gauge transformation \eqref{eq:simu1} labelled by $\mu(t)$. In comparison with 3D, it is simple to check that $\epsilon+\Lambda$ in 2D maps to $\epsilon$ in \eqref{diffasg}, and $\mu$ maps to $\eta$.

\subsection{Obstructions (and coincidences?)  in AdS$_2$}\label{sec:charge}

We now turn to building the conserved currents associated to the asymptotic symmetries constructed above.  For reasons that will be evident shortly, defining conserved charges and an algebra for them is not well defined in 2D gravity. The derivations below are intended to be suggestive, and the justification of this manipulations are motivated by the lift our boundary conditions to AdS$_3$ in section \ref{sec:pi}.

We start by defining the following currents
\bea\label{def:curr}
T^{ab}={2\over \ell^3\sqrt{-h}} {\delta I\over \delta h_{ab}}~,\quad J^{a}={1\over\ell \sqrt{-h}} {\delta I\over \delta A_{a}}~.
\eea
 Whereas $J_t$ will be the generator of $U(1)$ gauge transformations, actually $T_{tt}$ is not the generator of diffeomorphisms as one would naively expect. Under a diffeomorphism  $x^\mu \to x^\mu +\epsilon^\mu$ the action transforms as
\be\label{eq:NCD}
\delta_\epsilon I_{\rm 2D} = { \ell\over 2} \int dt \sqrt{-h}\,T^{ab}\delta_\epsilon g^{(0)}_{ab} +  \ell \int dt \sqrt{-h}\, J^{a}\delta_\epsilon A^{(0)}_{a}~.
\ee
Depending on the boundary conditions of the fields, if the term $J^{a}\delta_\epsilon A^{(0)}_{a} $ is sub-leading its contribution to the Noether current can be neglected (which sometimes justifies the naive identification of $T_{ab}$ as the generator of diffeomorphism). However, for the boundary conditions \eqref{bcond} this terms is {\it not} subleading relative to $T_{ab}$; there is a comparable contribution from the gauge field as we approach  $\sigma \to 0$.  Taking into account the contribution from the gauge field, we find that the {\it true} generator of diffeomorphisms is
\be\label{def:T}
\hat T _{tt}= T_{tt}+ J_t A_t~.
\ee

Our asymptotic symmetries are generated by a combination of diffemorphisms and gauge transformations. While $J_t$ does generate the $U(1)$ gauge transformations \eqref{eq:simu1}, the current $\hat T_{tt}$ is not the appropriate object that describes the combined transformation \eqref{2d:diffasg} and \eqref{gaugeasg}. We need to construct a current ${\cal T}$ which we can associate with  ``diff+gauge'' simultaneously. 

A way to derive  ${\cal T}$  is via  the Noether procedure, and this requires writing $\Lambda$---the gauge parameter---as a function of the diffeomorphism generator $\epsilon^\mu$. From  \eqref{2d:diffasg} and \eqref{gaugeasg}, we find that 
\bea\label{eq:ss}
\Lambda+A^{(2)}\zeta=-{\ell^2\over 8}  F^{\mu\nu} \nabla_\mu \epsilon_\nu +A_\mu  \epsilon^\mu + \cdots~,
\eea
where the dots denote subleading terms in the radial direction. This allows us to write the transformation law for the boundary gauge field as\footnote{To derive \eqref{eq:AEL} the following identities are useful: $$\nabla_\mu F^{\nu\lambda}=0~,\quad [\nabla_\mu,\nabla_\nu]\epsilon_{\lambda}=R^\alpha_{~\lambda \mu\nu}\epsilon_{\alpha}~,$$ and $$R_{\alpha\lambda \mu\nu}= {R\over 2}(g_{\alpha \mu}g_{\lambda\nu}-g_{\alpha \nu}g_{\lambda\mu}) ~.$$} 
\be\label{eq:AEL}
\delta_{\epsilon+\Lambda}A^{(0)}_\mu=-{\ell^2\over 8}  F^{\nu\lambda} \nabla_\nu (\nabla_\mu \epsilon_\lambda+\nabla_\lambda \epsilon_\mu)={\ell^2\over 8}  F^{\nu\lambda} \nabla_\nu \delta g_{\mu\lambda}~.
\ee
For those diffeomorphism $\epsilon^\mu$ that preserve the boundary conditions, \eqref{eq:AEL} will be zero as expected. If $\epsilon^\mu$ induces a change on the boundary metric $g^{(0)}$ we will assume that $A^{(0)}$ transforms according to \eqref{eq:AEL}.  The advantage  is that we can manipulate the infinitesimal transformations $\delta_\epsilon$ and $\delta_{\Lambda}$ in equal footing when evaluating variations of the action. The action under the combined diffeomorphism and gauge transformation is given by
\bea
\delta_{\epsilon+\Lambda} I &=& {1\over 2} \int dt \sqrt{-h}\, T^{ab}\delta_\epsilon g^{(0)}_{ab} +  \int dt \sqrt{-h}\, J^{a}\delta_{\epsilon+\Lambda}A^{(0)}_a \cr
&=& {1\over 2} \int dt \sqrt{-h}\, (T^{ab}-{\ell^2\over4}\nabla_c (J^aF^{c b}))\delta_\epsilon g_{ab}^{(0)}  ~.
\eea
From here we can read off  the stress tensor corresponding to this   ``diff+gauge'' transformation
\be \label{eq:gtrue}
{\cal T}(t) =T_{tt}- {\ell^2\over 4}\nabla_\sigma (J_t  F^{\sigma}_{\phantom{\sigma} t} )~.
\ee
 In our opinion, this derivation is unsatisfactory: we are mixing boundary and bulk covariant derivatives, and it is not clear how to carry out the integration by parts in our boundary integrals. There are as well other issues related to the proper definitions of charges in 2D which we discuss around \eqref{QQ}. It would be interesting to have a better derivation for ${\cal T}$.

Nevertheless, lets consider \eqref{eq:gtrue} and wonder if it has any interesting properties. To evaluate the stress tensor one needs to construct the boundary counterterm action. A detailed derivation can be found in e.g. \cite{Castro:2008ms}; the final answer for our setup is 
\bea\label{bla}
I_{\rm 2D,boundary}= {\ell\over  8 \pi G_2} \int dt \sqrt{-h} e^{-\psi} \left[K-{1\over 2\ell} + {\ell\over 2}e^{-2\psi}A_aA^a\right]~.
\eea
With this action principle, $I_{\rm 2D}=I_{\rm 2D,bulk}+I_{\rm 2D,boundary}$, we find that the on-shell value of the generators are 
\bea
J_t&=&{\ell\over  8 \pi G_2}e^{-3\psi^{(0)}}A^{(2)}  ~,\cr
{\cal T}&=& {\ell\over 16\pi G_2}e^{-3\psi^{(0)}}\left(  4 A^{(0)}  A^{(4)} - A^{(2)}A^{(2)} \right)~,
\label{calT}\eea
where we used \eqref{def:curr}, \eqref{def:T} and \eqref{eq:gtrue}, and we only kept the first non vanishing contribution in $\sigma$. Notice that there is a negative energy contribution to the stress tensor due to the `boundary' photons $A^{(2)}$ as we saw in \eqref{eq:3q} for the 3D charges.

The transformation properties of the generators $\cal T$ and $J_t$ with respect to the asymptotic symmetry group reads
\bea\label{eq:sc}
\left(\delta_{\epsilon +\Lambda}+\delta_ {\mu} \right){\cal T}&=& -\zeta \partial_t {\cal T}-2{\cal T}\partial_t \zeta  - J_t \partial_t \mu +{\ell e^{-\psi^{(0)}}\over 32\pi G_2}   \partial_t^3\zeta~,\cr
\left(\delta_{\epsilon +\Lambda} +\delta_{\mu} \right)J_{t}&=& -\zeta \partial_t J_{t}-J_{t}\partial_t \zeta+  {\ell e^{-3\psi^{(0)}} \over 8\pi G_2} \partial_t \mu~,
\eea
where we used \eqref{eq:he}, \eqref{eq:AAA} and \eqref{eq:simu1}. In addition to the expected tensorial transformation properties for ${\cal T}$ and $J_t$, we find an anomalous term for the stress tensor due to the combined diffeomorphism and gauge transformations,  and a $U(1)$ anomaly for the current.\footnote{Notice that $\left(\delta_{\epsilon +\Lambda}+\delta_ {\mu} \right){\hat T}_{tt} = -\zeta \partial_t {\hat T}_{tt}-2{\hat T}_{tt}\partial_t \zeta  - J_t \partial_t \mu$. The generator of diffeomorphism has no central extension as it should in a theory of AdS$_2$ gravity \cite{Strominger:1998yg}.}

One has to be rather careful in giving a physical interpretation to ${\cal T}$ and $J_t$. The dual theory to AdS$_2$ is  a quantum mechanical system, where the Hamiltonian is trivial and  the degeneracy of the ground state  is the non-trivial observable of this CFT$_1$. In the present discussion this is reflected in the conserved charge associated to ${\cal T}$; schematically the Noether charge behaves as
\be\label{QQ}
{\cal Q}_{\rm 2D} \sim  \sqrt{-h}\, {\cal T}^{t}\,_t \epsilon^t  \underset{\sigma \to 0 }\to 0~.
\ee
All charges associated to diffeomorphisms vanish as we take the UV cutoff $\sigma\to 0$, and hence all fluctuations encoded in ${\cal T}$  have vanishing energy. This is not surprising given the well known results in e.g. \cite{Maldacena:1998uz,Sen:2008yk,Amsel:2009ev,Dias:2009ex}. 

 Our discussion could simply end here: we could conclude that this theory describes solely the ground state and our asymptotic symmetry algebra is pure gauge.  But we will insist on giving an interpretation to ${\cal T}$ and $J_t$ in terms of a chiral (warped) CFT. 
We emphasize that from the AdS$_2$ perspective this is not well justified; the remaining of this section is just a report of some curious coincidences. 

If we take seriously \eqref{eq:sc}, we can identify the anomalous contributions as the central charge and $U(1)$ level of one copy of a Virasoro algebra plus a $U(1)$ Kac-Moody algebra;  normalizing the currents by the compactification radius, $\ell e^{-\psi^{(0)}}$, we have
\be\label{eq:ck2d}
c_{\rm 2D} = {3\over 4\pi G_2}  ~,\quad k_{\rm 2D}=  -{e^{-2\psi^{(0)}}\over 2\pi G_2}~.
\ee
This is consistent with  \eqref{eq:ck3} and \eqref{c:hs}, which are obtained independently. Regardless the choice of normalization, the ration between the anomalous terms in the currents $(c/k)$ in 2D are the same as those in  3D.

As a final remark, one could alternatively enforce boundary conditions where $A^{(2)}$ is fixed. Even though this choice is consistent, it is very restrictive on the classical phase space.  In particular setting $\delta A^{(2)}=0$  modifies the construction of the boundary counterterms in \eqref{bla} and we will end up with a trivial stress tensor.  $\delta A^{(2)}=0$ is the boundary condition used in \cite{Sen:2008yk,Sen:2008vm}, and in that case the theory is only capturing the degeneracy of states of the ground states. As we argued in section \ref{sec:glue} there are many cases where the reasonable (physical) boundary is  to fix $A^{(2)}$, while in other cases is more natural to choose  allow $A^{(2)}$ to fluctuate. This choice will depend on the global properties of the system under study.

\section{Improved bulk stress tensor: comparison with \cite{Hartman:2008dq}. }\label{app:hs}

In this appendix we will review the derivations done  in \cite{Hartman:2008dq}, and compare them with the results reported here.  In particular the improved bulk stress tensor used in \cite{Hartman:2008dq} does not have the canonical transformation properties and this modifies the relationship between the central charge and level. Following their analysis, we will show how to obtain the actual improved stress tensor which generates the asymptotic symmetries; these corrected expressions are in complete agreement with our results in the main text. 

Since the notation and technique used in \cite{Hartman:2008dq} is  different from ours, we first gather some of the key ingredients of their notation and analysis.  The action is
\be\label{hs:action}
S_{\rm HS}= {1\over 2\pi}\int d^2t \sqrt{-g}\left(\eta(R+{8\over \ell^2})-{2\over \ell^2}f^2 +f \epsilon^{\mu\nu}F_{\mu\nu}\right)~.
\ee
Here the scalar $\eta$ plays the same role as $e^{-\psi}$ in \eqref{2daction}, and $f$ is an auxiliary field used to eliminate the quadratic term in the gauge field $F_{\mu\nu}$. The values of the couplings in this action are not exactly the same as those used in the main text; this can be easily compensated by simple field redefinitions. 

The choice of gauge for the metric and gauge field are 
\be
ds^2=-e^{2\rho} dt^+ dt^-~,\quad t^{\pm}=t\pm \sigma~,
\ee
and
\be
A_{\pm}=\pm \partial_\pm a ~,\quad F_{+-}=-2\partial_-\partial_+ a~.
\ee
Note that this gauge differs from \eqref{gauge}; however this will not affect our final conclusion.

Define
\be
T_{\pm\pm}=-{2\pi\over \sqrt{-g}} {\delta S_{\rm HS}\over \delta g^{\pm\pm}}~, \quad j_{\pm}=-2\pi  {\delta S_{\rm HS}\over \delta A_{\mp}}~,
\ee
then the gauge and gravitational constraints that follow from the action \eqref{hs:action} are
\bea\label{eq:hsc}
j_-&=&-2\partial_- f=0~,\cr
j_+&=&2\partial_+ f=0~,\cr
T_{--}&=&-2\partial_-\eta \partial_-\rho +2\partial_-f\partial_- a +\partial_-^2\eta=0~,\cr
T_{++}&=&-2\partial_+\eta \partial_+\rho +2\partial_+f\partial_+ a +\partial_+^2\eta=0~,
\eea
where the last equality is the on-shell condition. %
The remaining equations of motion imply that these currents are holomorphically conserved, i.e.
\be
\partial_-j_+=\partial_+j_-=0~,\quad \partial_+ T_{--}=\partial_-T_{++}=0~.
\ee

The Dirac brackets at fixed $t^+$ are
\bea\label{eq:dirac}
[\partial_- \rho(s^-), \partial_-\eta(t^-) ]&=&\pi \partial_-\delta(s^--t^-)~,\cr
[\partial_- a(s^-), \partial_-f (t^-) ]&=&-\pi \partial_-\delta(s^--t^-)~.
\eea
From these brackets, and \eqref{eq:hsc}, we get 
\bea\label{eq:tjb}
[T_{--}(t^-),j_-(s^-)]&=&-2\pi\p_{-}\delta(t^--s^-)j_{-}(s^-)+2\pi \delta(t^--s^-) \p_-j_{-}(s^-)~,\cr
[T_{--}(t^-),T_{--}(s^-)]&=&-4\pi\p_{-}\delta(t^--s^-)T_{--}(s^-)+2\pi \delta(t^--s^-) \p_-T_{--}(s^-)~,
\eea
which give the expected tensorial transformations for a stress tensor and current. And as in \cite{Hartman:2008dq}, we assume that the Dirac brackets of two currents is anomalous:
\be\label{eq:jjb}
[j_{-}(t^-),j_-(s^-)]=-\pi k_{\rm 2D} \p_{-}\delta(t^--s^-)j_{-}(s^-)~.
\ee
(This assumption is justified by the analysis in section \ref{sec:charge}, which follows from  \cite{Castro:2008ms,Castro:2009jf}).  

The improved stress tensor  constructed in \cite{Hartman:2008dq} is
\be
 \mathcal{T}^{\rm HS}_{\pm\pm}=T_{\pm\pm}\pm {E\ell^2\over4}\p_\pm j_\pm~,
\ee
where $E$ is the value of the background electric field; alternatively we can write $E^2\ell^4=\bar \eta$ where $\bar \eta$ is the value of the constant dilaton. However this is not the appropriate stress tensor that  generates ``diff+gauge'': from  \eqref{eq:tjb} and \eqref{eq:jjb} it is simple to the that  $j_-$ does not transform under  $\mathcal{T}^{\rm HS}_{--}$ as a one form. In particular we find
\bea
[\mathcal{T}^{\rm HS}_{--}(t^-),j_-(s^-)]&=&-2\pi\p_{t^-}\delta(t^--s^-)j_{-}(s^-)+2\pi \delta(t^--s^-) \p_-j_{-}(s^-)\cr &&+{\pi E\ell^2\over4} k_{\rm 2D} \p_{-}^2\delta(t^--s^-)j_{-}(s^-)~.
\eea

From the definition of $j_{\pm}$ and the above brackets, we have
\be 
[j_-(t^-),a(s^-)]=-\delta(t^--s^-)~,
\ee
so adding a term proportional to $\p_\pm^2a$ can restore the tensorial properties of the stress tensor. The {\it correct} improved stress tensor  is
\bea 
\mathcal{T}_{\pm\pm}&=&T_{\pm\pm}+{E\ell^2\over4}(\pm\p_\pm j_\pm-\pi k_{\rm 2D}\p_\pm^2a)~,\\ 
\mathcal{T}_{+-}&=&{\pi E\ell^2k_{\rm 2D}\over4}\p_+\p_-a~.
\eea
This should be the analog of generator found in  \eqref{eq:gtrue}. The Dirac brackets are then given by
\bea [\mathcal{T}_{--}(t^-),j_-(s^-)]&=&-2\pi\p_{-}\delta(t^--s^-)j_{-}(s^-)+2\pi \delta(t^--s^-) \p_-j_{-}(s^-)~,\cr
\,[\mathcal{T}_{--}(t^-),\mathcal{T}_{--}(s^-)]&=&-4\pi\p_-\delta(t^--s^-)\mathcal{T}_{--}(s^-)+2\pi\delta(t^--s^-)\p_-\mathcal{T}_{--}(s^-)\cr
&&-{\pi k_{\rm 2D}E^2\ell^4\over16}\p_-^3\delta(t^--s^-)~.
\eea
Note the sign difference in the central term found in \cite{Hartman:2008dq} for ${\cal T}^{\rm HS}_{--}$. In perfect agreement with \eqref{eq:ck3} and \eqref{eq:ck2d}, we find that the central charge is 
\be \label{c:hs}
c_{\rm 2D}=-{3\over8}E^2\ell^4 k_{\rm 2D}=-{3\over2}\bar \eta \, k_{\rm 2D}~,
\ee 
instead of the positive relative sign reported in \cite{Hartman:2008dq}. It is important to note that the new off diagonal term $\mathcal{T}_{+-}$ will affect the definitions of charges, however it does not contribute to the anomalous term and  $c_{\rm 2D}$ is unaffected by it.

\end{document}